\documentclass[useAMS,usenatbib]{mn2e}
\usepackage{natbib, graphicx, times}
\usepackage{amsmath}
\usepackage{mathtools}
\usepackage{epstopdf}
\usepackage{amssymb}
\usepackage{extarrows}
\usepackage{color}
\usepackage{enumitem}
\usepackage{scrextend}
\newcommand{\apj}{ApJ}
\newcommand{\apjl}{ApJ}

\newcommand{\aap}{A \& A}
\newcommand{\aaps}{A \& AS}

\newcommand{\icarus}{Icarus}
\newcommand{\mnras}{MNRAS}

\newcommand{\nat}{Nature}
\newcommand{\pasj}{PASJ}

\newcommand{\Mjup}{M_{J}}
\newcommand{\Mdot}{\dot{M}}
\newcommand{\Tg}{T_\mathrm{growth}}
\newcommand{\Mex}{M_\mathrm{excess}}

\setlist{leftmargin=7.5mm}

\newcommand{\lsim}{\mathrel{\rlap{\lower4pt\hbox{\hskip1pt$\sim$}}
    \raise1pt\hbox{$<$}}}                
\newcommand{\gsim}{\mathrel{\rlap{\lower4pt\hbox{\hskip1pt$\sim$}}
    \raise1pt\hbox{$>$}}}                

\title[Growing planet-induced vortices]{Slowly-growing gap-opening planets trigger weaker vortices}

\author[Hammer, Kratter \& Lin]{Michael Hammer$^{1}$\thanks{E-mail: mhammer@as.arizona.edu},
Kaitlin M. Kratter$^{1}$
and Min-Kai Lin$^{1,2,3}$\\
$^{1}$ Steward Observatory, University of Arizona, Tucson, AZ 85721, USA \\
$^2$ Steward Theory Fellow \\
$^3$ Institute of Astronomy and Astrophysics, Academia Sinica, Taipei 10617, Taiwan
}

\begin{document}

\date{Accepted XXX. Received YYY; in original form ZZZ}

\pagerange{\pageref{firstpage}--\pageref{lastpage}} \pubyear{2016}

\maketitle

\label{firstpage}

\begin{abstract}
The presence of a giant planet in a low-viscosity disc can create a
gap edge in the disc's radial density profile sharp enough to excite
the Rossby Wave Instability. This instability may evolve into
dust-trapping vortices that might explain the ``banana-shaped"
features in recently observed asymmetric transition discs with inner
cavities. Previous hydrodynamical simulations of  planet-induced
vortices have neglected the timescale of hundreds to thousands of
orbits to  grow a massive planet to Jupiter-size. In this work, we study the
effect of a giant planet's runaway growth timescale on the lifetime
and characteristics of the resulting vortex. For two different planet
masses (1 and 5 Jupiter masses) and two different disc viscosities 
($\alpha$~=3$\times 10^{-4}$ and 3$\times10^{-5}$), we compare the
vortices induced by planets with several different growth timescales
between 10 and 4000 planet orbits. In general, we find that
slowly-growing planets create significantly weaker vortices with
lifetimes and surface densities reduced by more than $50\%$. For the 
higher disc viscosity, the longest growth timescales in our study 
inhibit vortex formation altogether. Additionally, slowly-growing
planets produce vortices that are up to twice as elongated, with
azimuthal extents well above $180^{\circ}$ in some cases. These
unique, elongated vortices likely create a distinct signature in the
dust observations that differentiates them from the more concentrated
vortices that correspond to planets with faster growth timescales. 
{Lastly,} we find that the low viscosities necessary for vortex formation likely
prevent planets from growing quickly enough to trigger the instability in self-consistent models.


\end{abstract}



\section{Introduction} \label{sec:intro}
Recent observations by the Atacama Large Millimeter Array (ALMA) have
uncovered transition discs with large asymmetries in the dust observed
at mm and sub-mm wavelengths. A variety of mechanisms -- including
vortices triggered by the Rossby Wave Instability (RWI) -- have been
proposed to explain these discs around stars such as Oph IRS 48
\citep{vanDerMarel13}, HD 142527 \citep{casassus13, fukagawa13}, as
well as SAO 206462 and SR 21 \citep{perez14}. 

A local minimum or maximum in a disc's radial potential vorticity or
`vortensity' profile can excite the RWI 
\citep{lovelace99, li00, li01}. Such vortensity extrema often coincide
with extrema in the disc's surface density profile, which can arise at
dead zone boundaries \citep[e.g.][]{regaly12, lyra12, miranda16} and gap-edges
created by giant planets  \cite[e.g.][]{li05}. 
In both cases, the RWI leads to exponential growth of non-axisymmetric
modes that develop into {a set of} vortices in the non-linear regime. In low mass
discs, these vortices will eventually merge to form a single
``banana-shaped"  anti-cyclonic vortex in less than 100 orbits. {Single}
vortices offer a natural explanation for dust asymmetries because they
can trap dust particles efficiently in a pressure maximum
\citep[e.g.][]{barge95, birnstiel13, lyra13, fu14b}.

Recent observations of the radial gas density profiles interior to the 
asymmetries have found that some of these regions are depleted,
favouring the latter idea that planet-induced vortices are responsible
for the asymmetric structures in some transition discs
\citep{vanDerMarel16}. In this case, a giant planet clears out a gap
in the disc, creating sharp peaks in the radial density profile at {each gap
edge}. If the planet is massive enough to steepen {a} gap edge beyond
a critical structure, it will trigger the RWI \citep{ono16}. 
This only happens in discs with low viscosities ($\alpha <
10^{-3}$). Otherwise, the viscous evolution of the disc will smooth
out the gap edge and prevent the instability from arising
\citep[e.g.][]{deValBorro07}.   

Several groups have conducted numerical simulations of planet-induced
vortices to study how well they can explain the recent observations of
dust asymmetries \citep{bae16} and to explore whether they can survive
for long enough to make them observable. In particular, \cite{fu14a}
used high-resolution 2D simulations to find that {a vortex induced 
by a 5 Jupiter-mass planet} can survive for up to 14000 orbits 
with optimal disc viscosities and temperatures. However, vortex lifetimes can
be weakened by a variety of different factors in particular regimes,
including self-gravity in very massive discs \citep{mkl11},
three-dimensional structure \citep{mkl12, meheut12}, and dust feedback
from large grains or high concentrations \citep{fu14b}. It remains an
open question as to how long vortices can survive and whether we
should expect them to be observable. 

One related question that has been relatively unexplored is the effect
of a planet's growth rate on the properties of the vortex. A planet's
growth rate can affect the evolution of the vortex it creates because
the planet changes the gap structure as it grows. Previous simulations
of planet-induced vortices have neglected realistic growth timescales
for the planet, only allowing it to reach its full mass in less than
100 orbits \citep[e.g.][]{deValBorro07, fu14a, loboGomes15, bae16}. 
In this regime, the vortex does not form until after the planet has
attained its final mass. With longer growth times, a Jupiter-mass
planet will trigger the RWI and induce vortices long before it
approaches its final mass \citep{mkl10}, giving the planet and the 
vortex plenty of time to restructure the gap edge and  impact the
vortex's further evolution. 

Models of core accretion -- the mechanism for forming most
Jupiter-mass planets -- show that a gas giant does not grow to full
size within a few dynamical timescales. The bulk of a planet's growth
occurs during the runaway gas accretion phase, when the mass of the
planet's gas envelope increases from a few Earth masses to a few
hundred Earth masses \citep{pollack96}. 
 Although this phase can be fast compared to the timescale needed for envelope accretion prior to runaway growth, it is limited by the depleted gas supply in
 the planet's vicinity once it opens up a gap \citep[e.g][]{lissauer09}. With
 the disc only gradually supplying the planet with gas, the runaway
 gas accretion phase can be slowed down to timescales of hundreds of
 orbits \citep{dangelo08}, if not thousands in lower viscosity discs
 \citep{lissauer09}. This is one or two orders of magnitude longer
 than the growth timescales used in earlier studies of planet-induced
 vortices.  

In this work, we explore the effect of a gap-opening planet's growth
timescale on the lifetimes and other properties of vortices induced by
the RWI. We organize this paper as follows: In
Section~\ref{sec:methods}, we describe our simulation set-up and
quantify our metrics for assessing vortex properties. In
Section~\ref{sec:results}, we present the results from our
simulations. In Section~\ref{sec:observations}, we discuss dust
concentration in elongated vortices, and put our simulation results in
the context of realistic planet formation models.   
In  
Section~\ref{sec:conclusions}, we summarize our results.  

%


\section{Methods} \label{sec:methods}

\subsection{Numerical Methods} \label{ssec:num_methods}

We use FARGO {\citep{FARGO}} to carry out simulations of a gas giant
planet with a prescribed growth rate embedded in a low viscosity
disc. FARGO is a 2D finite-difference hydrodynamical code that solves
the Navier-Stokes equations in cylindrical polar coordinates $(r,
\phi)$. It incorporates the FARGO algorithm, which allows for larger
integration timesteps by subtracting out the average azimuthal
velocity in each annulus when evaluating the Courant-Friedrich-Levy
condition. This technique makes it feasible to carry out a parameter
study of simulations with sufficient resolution beyond $10^4$ planet
orbits. 

We prescribe the growth of the planet's mass $m_p$ as a function of time $t$ as
\begin{equation} \label{eqn:growth}
m_p(t) = M_p\times \begin{cases}
\sin^2{\left(\pi t / 2T_\mathrm{growth}\right)}  & t\leq T_\mathrm{growth}, \\
1 & t >   T_\mathrm{growth},
\end{cases}
\end{equation} 
where $M_p=m_p(T_\mathrm{growth})$ is the planet's final mass and
$T_\mathrm{growth}$ is the planet's growth timescale, the key
parameter in our study.  

In principle, one can calculate a ``self-consistent" growth rate based
on the amount of material within the planet's Roche radius. However,
we instead use the fixed growth prescription above because it allows
us to conduct a controlled parameter study of growth time alone.  
At the resolution of our fiducial simulations, the circumplanetary
discs are not sufficiently well resolved to generate realistic growth
rates \citep{zhu15}. Moreover, accretion onto the planet is a function
of a more complex set of disc parameters dependent on thermodynamics
and feedback \citep{ayliffe09}, which we neglect in this study. 
To avoid an unphysical build-up of material within the planet's Roche
radius, we remove mass from the computational domain according to the
prescription from \citet{kley99} that is implemented for FARGO's
accretion scheme. {The total amount of mass removed is relatively small and never exceeds $5 \%$ of the planet's mass.}

At the radial boundaries, we apply wave killing zones
\citep[e.g.][]{deValBorro06} in $r \in [1,1.25]r_\mathrm{in}$ and in
$r\in[0.84,1]r_\mathrm{out}$, where $r_\mathrm{in}$ and
$r_\mathrm{out}$ are the inner and outer edges of the disc
respectively. The outer zone is rapidly damped on a timescale of
1/500th of the orbital period at the outer boundary so that it
approximates a mostly unperturbed region. The inner region is damped
at a softer rate of 1/3rd of the innermost orbital period. Periodic
boundaries are applied in the azimuth.

\subsection{Simulations} \label{ssec:simulations}

We initialize each simulation with a locally isothermal disc that has
a power law radial surface density profile of $\Sigma(r)  = \Sigma_0
(r/r_p)^{-1}$, where $\Sigma_0$ is the initial surface density at the
fixed orbital radius of the planet, $r_p$. The value of $\Sigma_0$ is
set by the total disc mass in the domain, which we fix to $M_{d} =
2M_p$. We do not include self-gravity because the disc mass is very low with a Toomre $Q > 20$ at the location of the vortex in all of our simulations \citep{toomre64}. 
The temperature profile is set to
fix the disc aspect ratio $h\equiv H/r= 0.06$, where $H$ is the disc's
scale height. We selected this value of $h$ because \cite{fu14a} have
shown it maximizes vortex lifetimes.

We simulate an annulus of a disc across a radial domain that extends
from $r\in[0.2,5.7]r_p$ and $\phi\in[0,2\pi]$ in azimuth. This {annulus} is
resolved by $N_r \times N_\phi = 1024 \times 2048$
arithmetically-spaced grid cells in the radial and azimuthal
directions respectively. This grid spacing resolves both the 
disc's scale height $H(r)$ and the planet's Hill radius by at least 11
cells in the the outer disc ($r > r_p$), our region of interest. Our
standard resolution in this region is slightly higher than those used
in most other recent studies of planet-induced vortices 
\citep[e.g.][]{loboGomes15, les15, bae16}, but less than the highest
resolution studies 
 \citep{fu14a}. We carry out convergence tests for select cases at a 
 lower resolution of $768 \times 1536$ and a higher resolution of
 $1536 \times 3072$ (see Section~\ref{ssec:resolution}). 

In the disc, we place a planet on a fixed circular orbit around a
solar mass star ($M_*=M_\odot$) with an orbital frequency $\Omega_p =
\sqrt{GM_\odot/r_p^3}$ that corresponds to an orbital period of $T_p
\equiv 2\pi/\Omega_p$, where $G$ is the gravitational constant. Our
computational units are such that $r_p=\Omega_p=G=M_*=1$. The planet's
gravitational potential is smoothed out to $0.6r_H$, where $r_H
=(q/3)^{1/3}r_p$ is the planet's Hill radius and $q$ is the 
planet-to-star mass ratio. In our parameter study, we use two
different final planet masses of $q=1 \times 10^{-3}$ and
$5\times10^{-3}$, which correspond to $M_p = 1$ and $5~M_J$ (Jupiter
masses) for a solar mass star. We also use two different disc
viscosities of $\nu = ~10^{-6}$ and $10^{-7}$ in units of $r_p^2
\Omega_p$, corresponding to $\alpha \approx 3 \times 10^{-4}$ and $3
\times 10^{-5}$ respectively for the standard alpha prescription of
$\nu = \alpha H^2\Omega_p$ \citep{alpha}. For each of these 4
combinations of planet masses and disc viscosities, we explore the
evolution of the RWI with a focus on the properties and lifetimes of
the vortices that form for four different planet growth timescales
(see Table~\ref{table:simulations}). 


\begin{table}
\caption{Planet growth times for each of the four cases in our simulations.}
\begin{tabular}{ c | c | l }
  $M_p~/~M_J$ & $\nu~/~r_p^2 \Omega_p$ & $T_{\rm growth}~/~T_p$ \\
   \hline
  1 & $10^{-6}$ & 10, 250, 500, 1000 \\
  1 & $10^{-7}$ & 10, 500, 1000, 2000 \\
  5 & $10^{-6}$ & 10, 500, 1000, 2000 \\
  5 & $10^{-7}$ & 10, 1000, 2000, 4000 \\
\end{tabular}
\label{table:simulations}
\end{table}

\subsection{Identifying a Vortex Signature} \label{ssec:identify}

To quantify the vortex properties as a function of the planet's growth
time ($T_\mathrm{growth}$), we need a robust method to identify when
they form and dissipate. Quantitative definitions are especially
important since {vortices do not transition from `nonexistent' to `newly-formed,' or from `weak' to `dead' instantaneously.}



We find that the magnitude of the overdensity within the vortex is the
best tracer of formation and lifetime. While the planet also produces
density enhancements in its spiral wake, these remain fixed in the 
planet's co-rotating frame. On the contrary, a vortex orbits at its
local Keplerian frequency several Hill radii away from the
planet. This difference allows us to differentiate the vortex
over-density from the spiral density waves. We isolate the vortex
(with its distinct Keplerian orbital frequency) from the spiral
density waves (co-rotating with  the planet) by differencing the
surface density fields at successive planet orbits when the planet is
at the same location. We find this procedure to be effective in
removing the co-rotational features, leaving only the vortices present
in the positive component of the differenced surface density. We then
combine all of the positive contributions and define this mass to be
the total mass in the vortex, $M_{\rm excess}$. We use this ``excess
mass" to quantitatively determine when a planet triggers the RWI and
when a vortex dissipates.  

We numerically identify when the RWI is first triggered as the first
orbit at which there is a sharp increase in the growth rate of $\Mex$,
that is, when it begins to increase roughly exponentially. These times are
straightforward to identify from the evolution of the growth rates
over time. For a given disc viscosity $\nu$, we find that these times
correspond to a very narrow range of trigger masses across our
parameter space. As a result, we do not need a more precise
quantitative definition of the trigger point.  

While vortices are easy to identify at their peak intensity,
pinpointing the orbit at which they disappear is challenging. Vortices
can persist at very low mean densities for hundreds of orbits. Thus,
we quantify their lifetime based on when {$M_{\rm excess}< 0.20~\Sigma_0 r_p^2$ because at this mass, we find that the more concentrated vortices in our study have typically lost $\approx 85-95\%$ of their peak excess masses (which fall in the range of 1.5 to 3.5 $\Sigma_0 r_p^2$), leaving them with an average overdensity of no more than $10\%$ above the background density.}
{Meanwhile, the choice of a threshold mass does not have a significant effect on the lifetimes of less concentrated vortices since they dissipate more abruptly.}


Although it might seem natural to include a quantitative measure of
vorticity in our definition of vortex formation and dissipation, we
found this to be less reliable. This is because other features at the
gap edge also produce spikes in vorticity that make it more difficult
to differentiate the vortex's vorticity from that of nearby parts of
the disc. 

\subsection{Characterizing Vortex Strength} \label{ssec:strength}

We characterize a vortex's strength by quantifying its surface density
and azimuthal extent. In general, stronger long-lived vortices have
higher surface densities and more concentrated azimuthal extents. 

We measure the maximum gas concentration a vortex attains during its
lifetime through its peak surface density,
$\Sigma_\mathrm{peak}$. This is the maximum value of $\Sigma(r =
r_\mathrm{vortex}, \phi = \phi_\mathrm{peak}, t)$ over time, where
$r_\mathrm{vortex}$ {corresponds to the orbital radius of the vortex defined as the} radius at which there is a local
maximum in the outer disc ($r > r_p$) in the azimuthally-averaged
surface density profile; 
$\phi_\mathrm{peak}$ is the azimuth at which
$\mathrm{max}\left[\Sigma(r = r_\mathrm{vortex},\phi)\right]$
occurs. We apply a smoothing filter over 5 orbits before maximizing
these values over time. This removes short timescale variations, such
as those {to} due the planet's wake. We choose to use the vortex's maximum
surface density rather than its average surface density to avoid the
difficulty of tracing out the edges of the vortex, which can be
ambiguous in many snapshots. Nonetheless, we estimate that the
vortex's average surface density is smaller than its maximum surface
density by $25$ to $35\%$. 

We estimate a vortex's azimuthal extent, denoted as $\Delta \phi$, by
examining $\Sigma(r = r_\mathrm{vortex},\phi)$ over a wide range of
snapshots. We find that a vortex's azimuthal extent is the most
concentrated near the time at which its surface density is 
maximized. For this reason, we characterize a vortex's concentration
by the minimum value of $\Delta \phi$ over time, denoted as
$\left[\Delta \phi \right]_\mathrm{min}$. {Each concentration is} rounded to the nearest
$30^{\circ}$, and primarily intended to classify a vortex as
``concentrated" ($\left[\Delta \phi \right]_\mathrm{min} <
180^{\circ}$), ``elongated" ($\left[\Delta \phi \right]_\mathrm{min} >
180^{\circ}$), or ``intermediate" ($\left[\Delta \phi
  \right]_\mathrm{min} \approx 180^{\circ}$). Section~\ref{ssec:dust}
also uses these rough values to show how vortices with different gas
azimuthal extents appear in the associated dust observations.

\section{Results} \label{sec:results}


\begin{figure*} 
\centering
\includegraphics[width=0.47\textwidth]{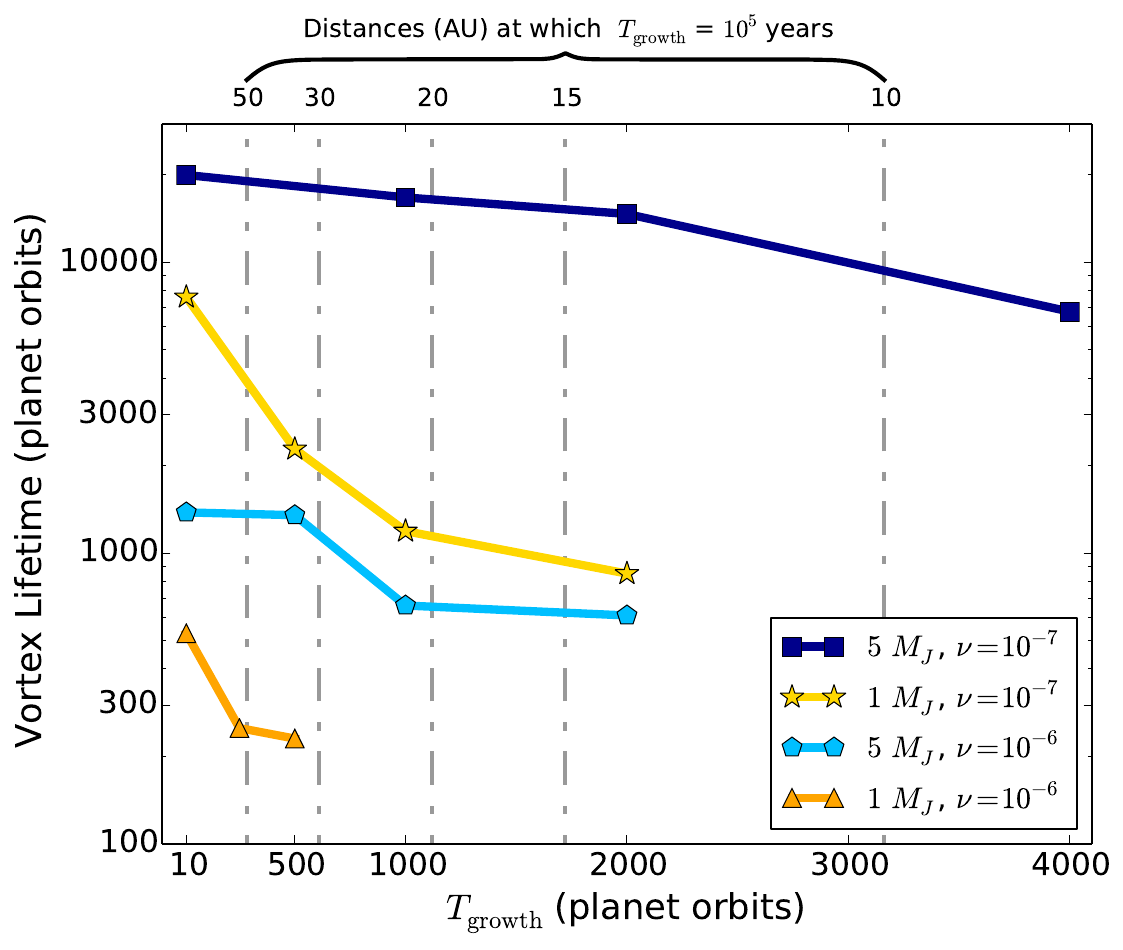}
\hspace*{\fill}
\includegraphics[width=0.455\textwidth]{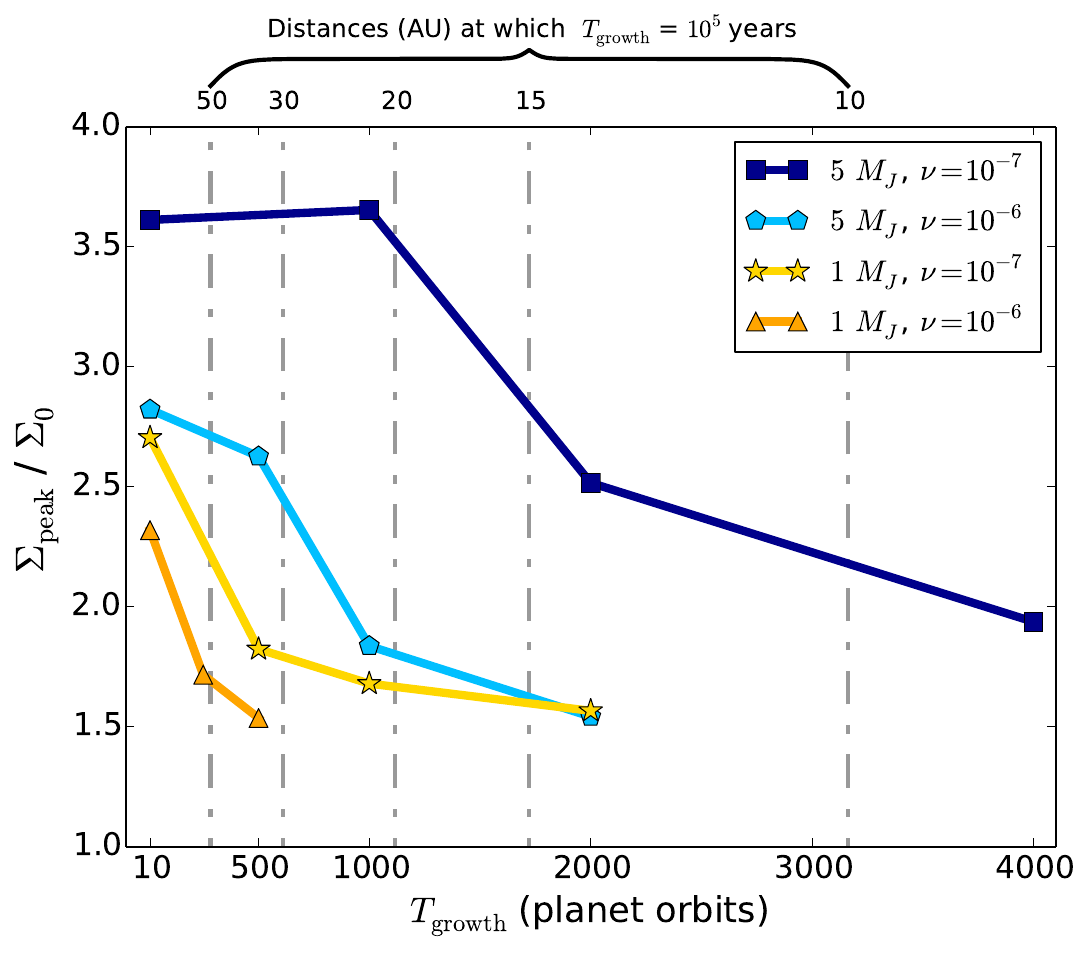}
\caption{Vortex lifetimes (left panel; defined in
  Section~\ref{ssec:identify}) and peak surface densities
  $\Sigma_\mathrm{peak}$ (right panel; defined in
  Section~\ref{ssec:strength}) as a function of planet growth
  timescale $\Tg$ for each of the four cases in our study. Each dashed
  vertical line show the minimum distance at which a growth timescale
  in orbits implies a growth time of at least $10^5$ years (with
  $M_{*} = 1~M_{\odot}$ assumed), which serves as a benchmark
  timescale from circumplanetary disc accretion rates (see
  Section~\ref{ssec:constraints}).  
Left: Vortex lifetimes decrease monotonically with larger $\Tg$ in
each case. Lower viscosity discs also yield longer vortex lifetimes. 
Right: $\Sigma_\mathrm{peak}$ essentially also decreases monotonically
with larger $\Tg$. Higher mass planets induce vortices with higher
$\Sigma_\mathrm{peak}$. The surface  densities are normalized by
$\Sigma_0$, the initial surface density at the planet's orbital
radius, which is also roughly the background density near the vortex.} 
\label{fig:comparisons}
\end{figure*}

Our simulations show that planets with longer, more realistic growth times induce vortices with:
\begin{enumerate}
  \item{ shorter  lifetimes,}
  \item  {lower  peak surface densities, and  }
  \item  {wider  azimuthal extents.}
\end{enumerate}
We summarize these trends in Figure \ref{fig:comparisons}. 
We can interpret these results by first considering the gap-opening process, which begins before the RWI is triggered.
According to \cite{crida06}, planets open gaps when
\begin{equation}
\label{eq:crida}
\frac{3}{4}\frac{H(r_p)}{R_H}+\frac{50}{q\mathrm{Re}} \lsim 1,
\end{equation}
where the Reynolds number $\mathrm{Re} = r_p^2\Omega_p/\nu$ is the
reciprocal of our dimensionless viscosity. Equation \ref{eq:crida}
implies gap opening masses of $M_\mathrm{gap} \approx 0.05~\Mjup$ and
$0.15~\Mjup$ for $\nu=10^{-7}$ and $10^{-6}$ respectively. These are
well below the final planet masses in our simulations.  
This suggests that the planet can trigger vortices through the RWI
\emph{during} its growth phase when its mass is still relatively
small.  This is distinct from previous simulations of planets of at
least Jupiter-mass where the planet reaches its full mass almost
immediately \emph{after} vortices form \citep[e.g.][]{deValBorro07,
  fu14a}.  

\begin{figure*} 
\centering

\includegraphics[width=0.70\textwidth]{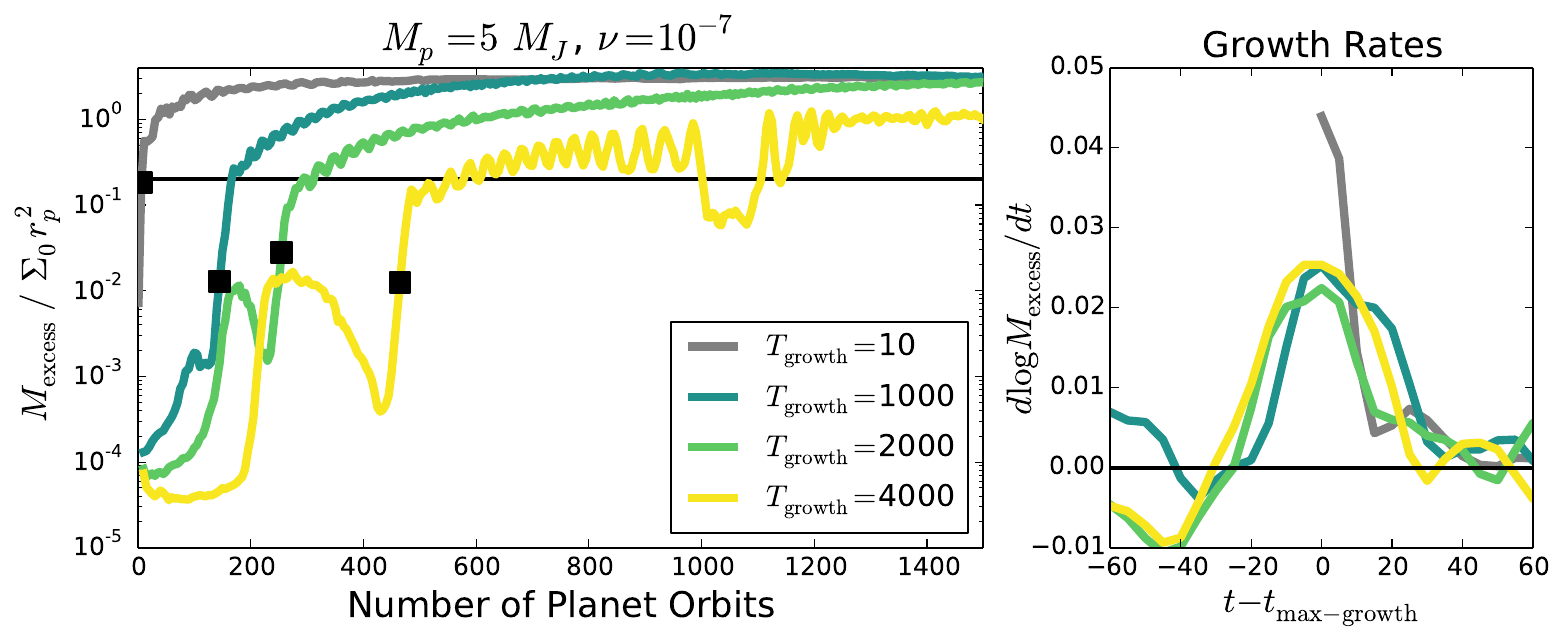}
\vspace*{2em}

\includegraphics[width=0.70\textwidth]{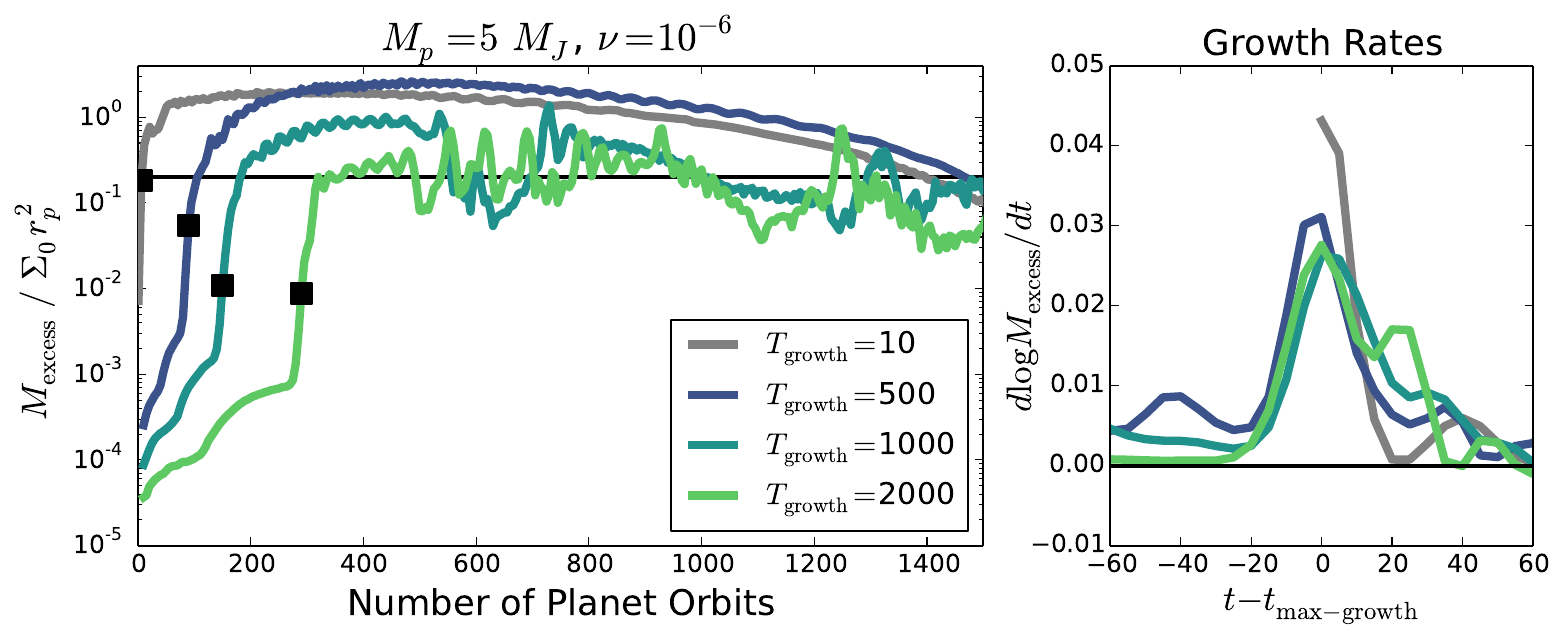}
\vspace*{2em}

\includegraphics[width=0.70\textwidth]{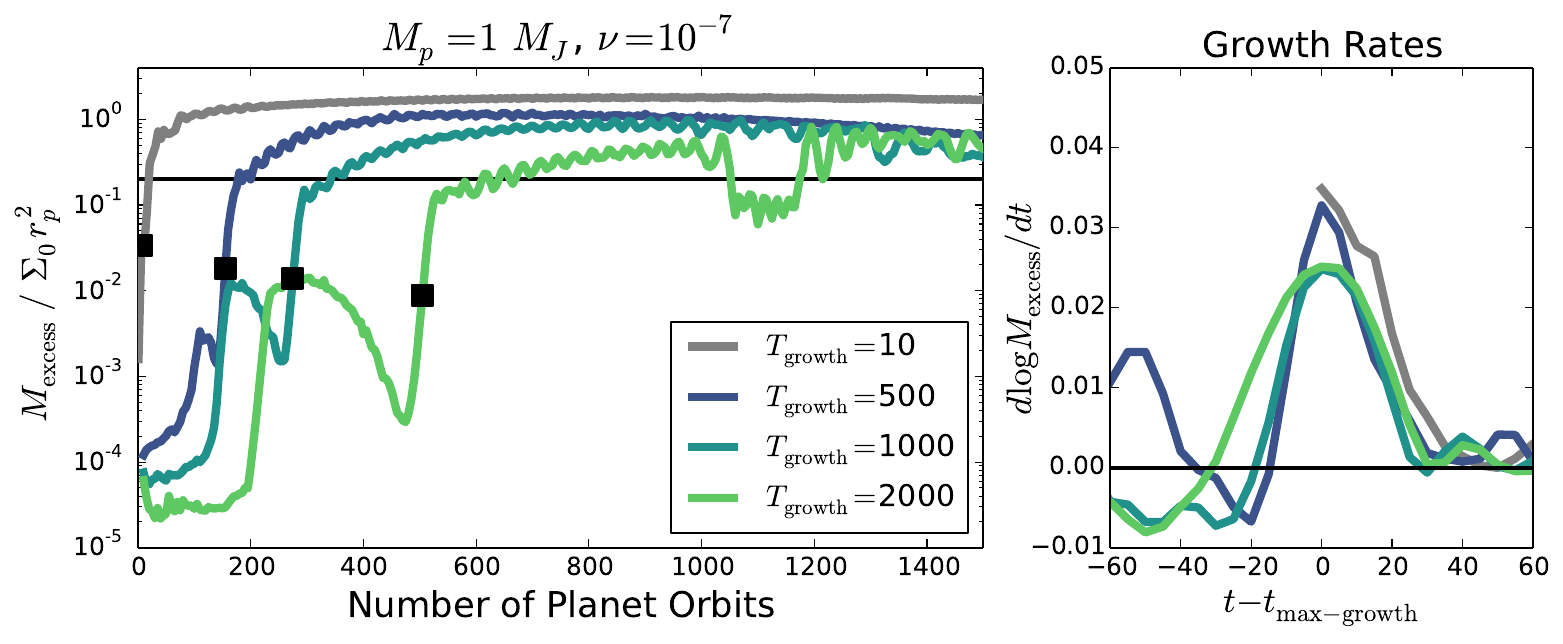}

\caption{Left panels: The excess mass, $M_{\rm excess}$, in a vortex
  over time across our parameter space. When $M_{\rm excess} >
  0.2~\Sigma_0r_p^2$ (marked with a horizontal line), a vortex is
  alive. Right panels: Exponential growth rates across our parameter
  space, during the early growth stages of each vortex and centred at
  the snapshot with the maximum growth rate (denoted in the left
  panels with black squares). Regardless of $m_p$ and
  $\nu$, all of the slowly-grown planets with $T_{\rm
    growth} \gg 10~T_p$ trigger vortices with similar maximum growth
  rates of $d\log M_{\rm excess}/dt \approx 0.025$. After reaching this
  point, the growth rates immediately begin to decline. 
}
\label{fig:growth}
\end{figure*}

Indeed, using $M_{\rm excess}$ shown in Figure \ref{fig:growth} ---
our measure of vortex presence --- we find that the first spike in
$M_{\rm excess}$ occurs when $m_p\approx 0.03-0.06~\Mjup$ for
$\nu=10^{-7}$, and at roughly $m_p \approx 0.2~\Mjup$ for
$\nu=10^{-6}$, both independent of $T_\mathrm{growth}$. 
These values of $m_p$ are close to the gap-opening masses
$M_\mathrm{gap}$ estimated above, suggesting that the RWI develops
soon after gap-opening begins. 
Once the RWI is triggered, $M_{\rm excess}$ enters a short exponential
growth phase lasting for $50-100$ orbits. We find this timescale,
$T_{\rm exp}$, is also approximately the time it takes for  
the initial vortices to merge into a single vortex (azimuthal
wavenumber $m=1$), and that $T_\mathrm{exp}$ does not vary
significantly with $\Tg$.   

Despite similar trigger masses and exponential growth phases, the
resulting vortices are in fact quite different {when the planet takes longer to grow}.
{These differences arise}
because once the vortex emerges, it begins to back-react on the gap
edge, smoothing it out. The growing planet competes against the
already-present vortex to sharpen the gap edge. \cite{varniere04} show
that the ratio of surface density at the gap centre,
$\Sigma_\mathrm{gap}$ to that at the gap edge, $\Sigma_\mathrm{edge}$,
depends on planet mass and Reynolds numbers as: 
\begin{equation}
\frac{\Sigma_\mathrm{edge}}{\Sigma_\mathrm{gap}} \propto q^2~\mathrm{Re}. 
\end{equation}
For a growing planet, we expect the steepness of the gap edge to
evolve on a timescale of order $T_\mathrm{gap} \equiv q(t) /
\dot{q}(t)$. 

\begin{figure} 
\centering
\includegraphics[width=0.47\textwidth]{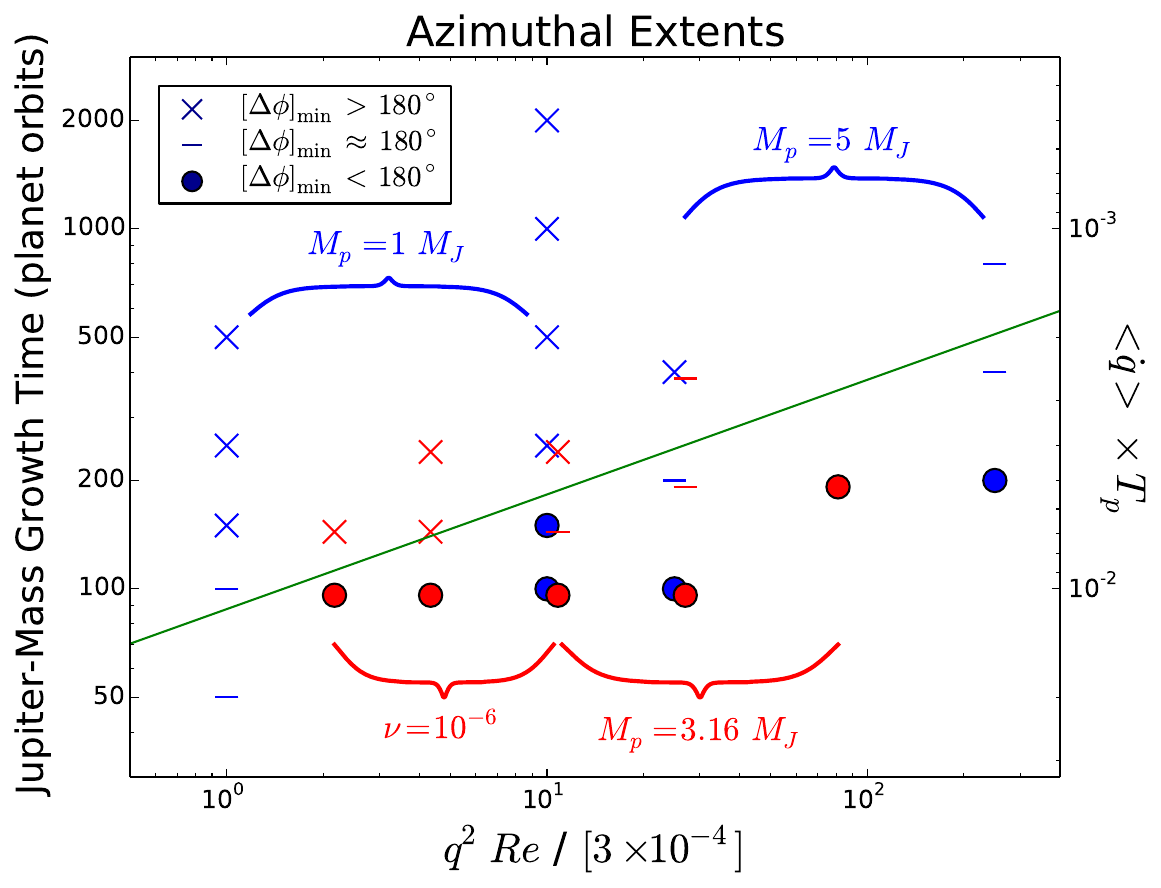}
\caption{Vortex shapes across our parameter space {(blue points are part of our main study; red points are added)}. 
The green line {shows that the two parameters on the axes related to gap steepening
  separate} elongated vortices (above) from concentrated ones
  (below). We determine the parameters for this dividing line by
  assuming a power law parametrization of $<\dot{q}~/~10^{-3}>^{-1} =
  A \times (q^2Re)^B$. We then use a soft-margin support vector
  machine model with $c = 25$ \citep[implemented by
    SVM-Light,][]{joachims99} to determine that the values of $A = 88$
  and $B = 0.32$ create the maximum margin between the set of
  elongated vortices and the set of concentrated vortices.} 
\label{fig:extents}
\end{figure}

In Figure \ref{fig:extents}, we show that the {azimuthal extents of the
vortices are well-divided} in a parameter space $q^2 \mathrm{Re}$
vs. $<\dot{q}>^{-1}$, where angle brackets represent a time average. Specifically, the properties of the vortex
depend on both the steepness of the gap edge and how quickly the gap
edge evolves. This dependence reflects the competition between $T_{\rm
  gap}$, the timescale for gap sculpting; and $T_{\rm exp}$, the
roughly constant timescale which governs how quickly the vortex can
back-react on the gap. To verify the robustness of our choice of
dimensionless numbers to map vortex properties, we ran short-term
simulations with different combinations of $q$ and $\nu$ that more
densely sample $q^2 \mathrm{Re}$, and overlap with values already
in our full length parameter study. We find excellent consistency between vortex
azimuthal extents in the overlapping cases. When $T_{\rm gap} > T_{\rm
  exp}$ (that is, for slow planet growth), the vortex begins to smooth
out the gap edge while the planet is still at a low mass, preventing the gap edge from attaining the
maximum steepness for a given pair of $q$ and $Re$. This leads to the vortex having a smaller
saturation amplitude in density and a less concentrated azimuthal
extent. When $T_{\rm gap} \lsim T_{\rm exp}$, the planet grows
quickly, continually steepening the gap edge. Thus, the back-reaction
from the vortex only becomes important when the planet is more massive and the gap edge
sharper. {This leads to the vortices becoming} stronger and more concentrated. 

For more massive planets with larger $q^2~\mathrm{Re}$, we observe
that the azimuthal extent of a vortex can evolve from ``elongated" to
``intermediate." This behavior arises because the planet continues
to steepen the gap edge long after the exponential growth phase of
$M_\mathrm{excess}$. As a result, the slope of the dividing line in
Figure \ref{fig:extents} is positive. We emphasize that the largest
distinction between the vortices is their azimuthal
extents, rather than their peak mass or saturation amplitude. This is
consistent with the linear scaling between the RWI growth rate and saturation
amplitude found by \cite{meheut13}.

With this interpretation in hand, we now provide more detailed
descriptions of the different cases. We describe the vortices induced
by $5~M_J$ planets in Section~\ref{ssec:highMass} and the vortices in
induced by $1~M_J$ planets in Section~\ref{ssec:lowMass}. We observe
and explain repeated vortex formation in
Section~\ref{ssec:reformation}. Lastly, we show that the underlying
trends in our results converge at different resolutions in
Section~\ref{ssec:resolution}.  

\subsection{Vortices induced by Higher Mass Planets ($M_p = 5~M_J$)} \label{ssec:highMass}


\begin{figure*} 
\centering

\includegraphics[width=0.98\textwidth]{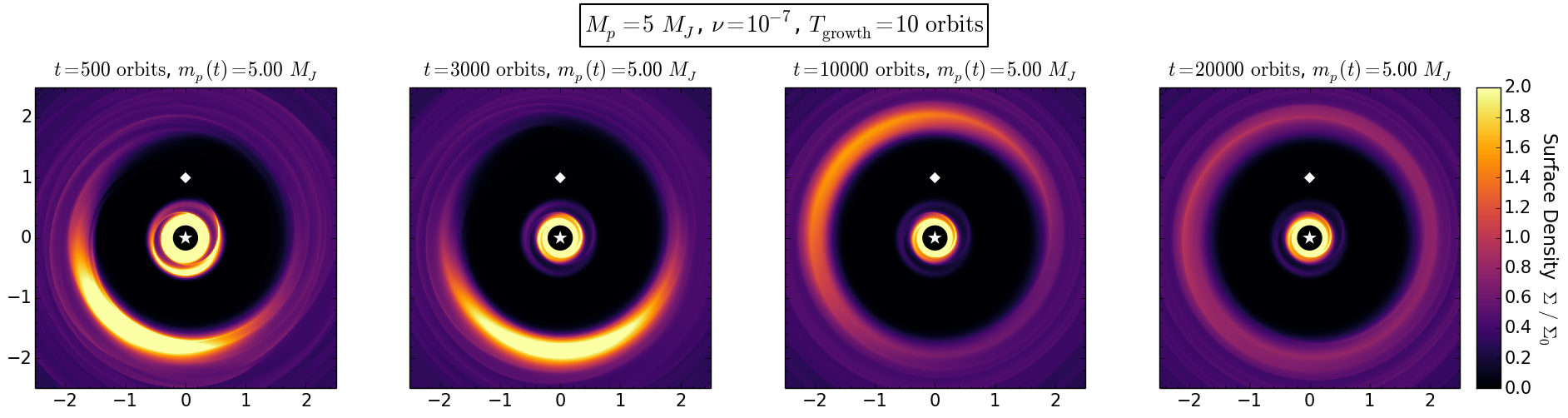}
\vspace*{2em}


\includegraphics[width=0.98\textwidth]{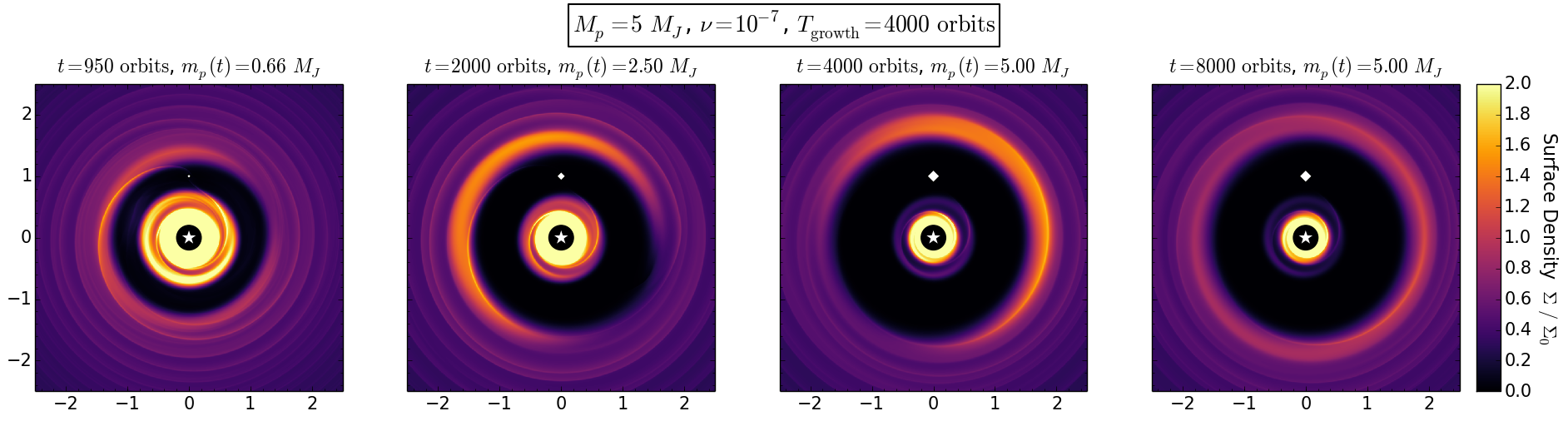}

\caption{Surface density snapshots for $M_p = 5~M_J$ and $\nu =
  10^{-7}$, for $T_{\rm growth} = 10$, $2000$, and $4000~T_p$. The
  title of each snapshot indicates the time, and the mass of the
  planet at that time. The planet is indicated by a diamond above the
  star in each snapshot, {and} its size is scaled relative to the planet's
  final mass.  
Top: With $T_{\rm growth} = 10~T_p$, a strong vortex forms very
quickly and persists at peak strength through 3000 orbits. After 10000
orbits, the vortex has weakened and spreads out slightly in azimuth,
but is still rather prominent. It does not die until $20000$ orbits.  
Despite starting out weaker, the vortices with $T_{\rm growth} =
1000$ and $2000~T_p$ {(not shown)} also remain prominent after $10^4$ orbits, just like
in the rapid growth case, {but ultimately dissipate earlier.}
Bottom: With $T_{\rm growth} = 4000~T_p$, the vortex's peak density
reaches only about half that of the rapid growth case, and never
becomes more concentrated than $180^{\circ}$. In-between the first two
snapshots shown, the vortex dies at about 1000 orbits and reforms
$\sim 50$ orbits later. After the planet finishes growing at 4000 orbits,
the vortex is only as strong as the vortices near 10000 orbits for the other
growth timescales {(compare the third column of panels)}. The vortex then fades faster, dying out within 8000
orbits.} 
\label{fig:pretty_m5_v7}
\end{figure*}


\begin{figure*} 
\centering

\includegraphics[width=0.98\textwidth]{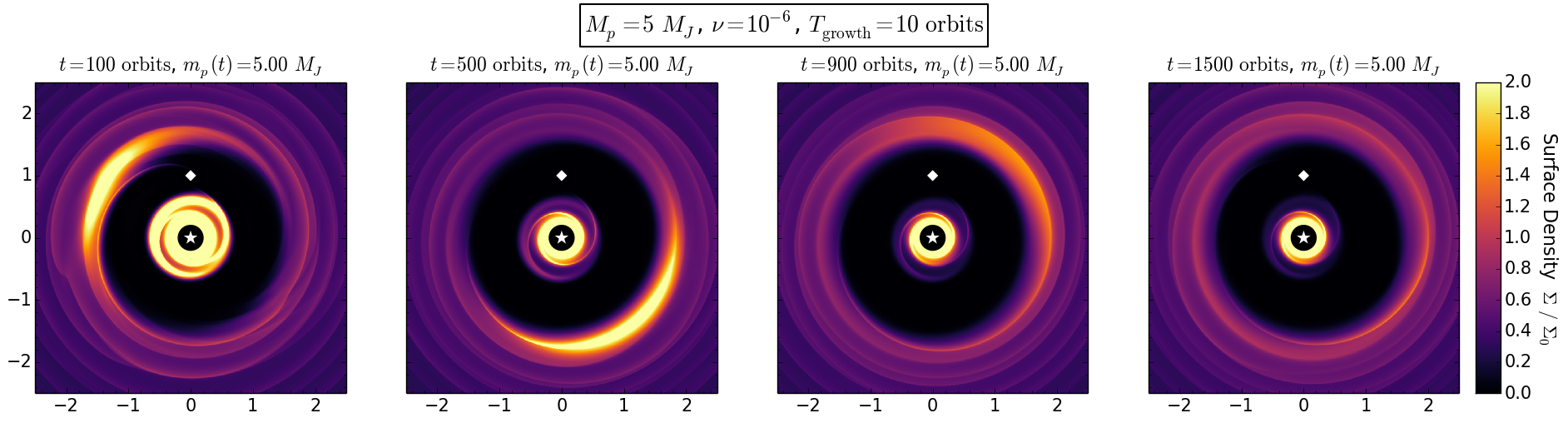}
\vspace*{2em}


\includegraphics[width=0.98\textwidth]{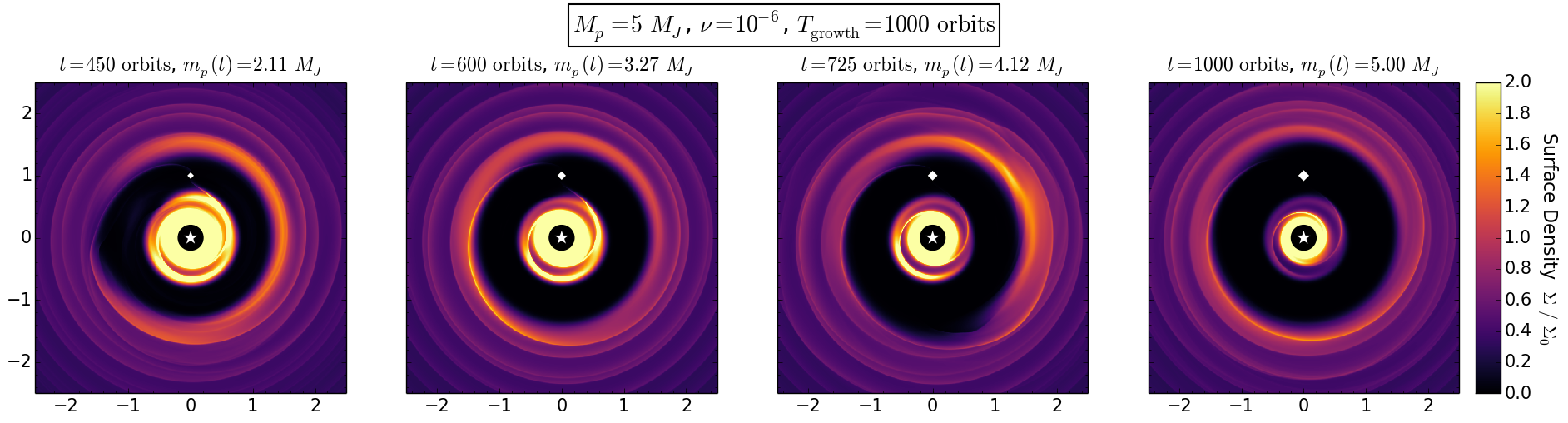}

\caption{Surface density snapshots for $M_p = 5~M_J$ and $\nu =
  10^{-6}$, for $T_{\rm growth} = 10$, $500$, and $1000~T_p$. See
  Figure~\ref{fig:pretty_m5_v7} for more details.  
Top: With $T_{\rm growth} = 10~T_p$, a strong vortex forms by 100
orbits and remains near peak strength at 500 orbits. By 900 orbits,
the vortex has quickly faded and by 1500 orbits, it has died out. 
The vortex with $T_{\rm growth} = 500~T_p$ {(not shown)} develops similar properties
to the vortex in this rapid growth case, despite forming slightly
later. 
Bottom: With $T_{\rm growth} = 1000~T_p$, a much weaker, more elongated
vortex initially forms. It dies out by 600 orbits. As the planet
continues to grow, it then re-triggers {a vortex of similar strength} about 100 orbits
later. {This} vortex then dies out near 1000 orbits when the planet has
finished growing. The longest timescale $T_{\rm growth} = 2000~T_p$ {(not shown)}
never produces a clear $m = 1$ vortex.}  
\label{fig:pretty_m5_v6}
\end{figure*}

We find that for $T_\mathrm{growth}=10~T_p$, planets with $M_p = 5~M_J$ in
discs with a low viscosity ($\nu = 10^{-7}$) generate strong,
concentrated vortices that survive well beyond $10^4$ orbits, in agreement
with previous studies \citep{fu14a}.
Unlike in any of the other cases (see below), planets with $T_\mathrm{growth}=2000~T_p$ also 
generate rather strong vortices with lifetimes just above $10^4~T_p$ and
$\Sigma_\mathrm{peak}>2.5~\Sigma_0$. Nevertheless, these vortices are
still weaker than those produced by a rapidly-grown planet. We find
that it takes increasing $T_\mathrm{growth}$ to $4000~T_p$ to reduce
the vortex's $\Sigma_\mathrm{peak}$ below $2~\Sigma_0$ (see 
Figure~\ref{fig:pretty_m5_v7}). These vortices still survive more than
7000 orbits with azimuthal extents that are relatively concentrated
($\left[\Delta \phi\right]_\mathrm{min}\approx 180^{\circ}$) rather
than elongated. 

In a high viscosity disc ($\nu = 10^{-6}$), a $5~M_J$ planet with $\Tg
= 500~T_p$ induces a vortex that survives just as long ($\approx 1300$
orbits) and has a similar surface density
($\Sigma_\mathrm{peak}\approx 2.7~\Sigma_0$) as a vortex generated by
a rapidly-grown planet. However, with $T_\mathrm{growth}=1000~T_p$,
the vortex's lifetime is halved, its azimuthal extent doubles (from
$\left[\Delta \phi\right]_\mathrm{min} = 90^{\circ}$ to
$180^{\circ}$), and 
$\Sigma_\mathrm{peak}<2~\Sigma_0$. Figure~\ref{fig:pretty_m5_v6} shows
snapshots of this case compared to the rapid growth case. With
$T_\mathrm{growth}=2000~T_p$, even though the RWI is triggered with
high-$m$ azimuthal modes, we do not observe them to evolve and merge into a
clear $m =1$ vortex.   

\subsection{Vortices induced by Lower Mass Planets ($m_p = 1~M_J$)} \label{ssec:lowMass}


\begin{figure*} 
\centering
\includegraphics[width=0.98\textwidth]{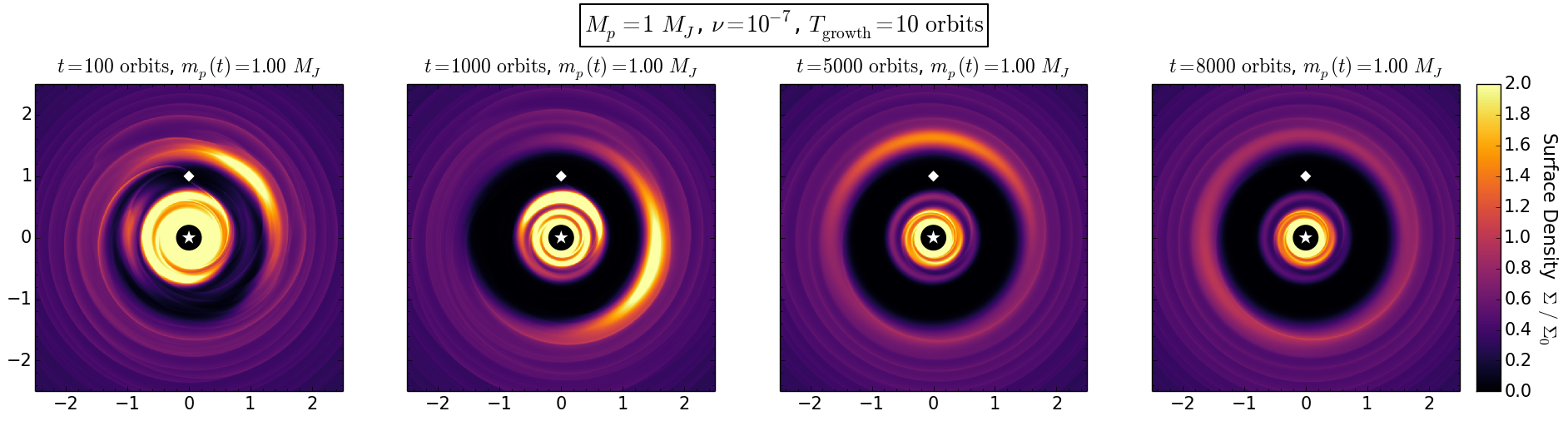}
\vspace*{2em}

\includegraphics[width=0.98\textwidth]{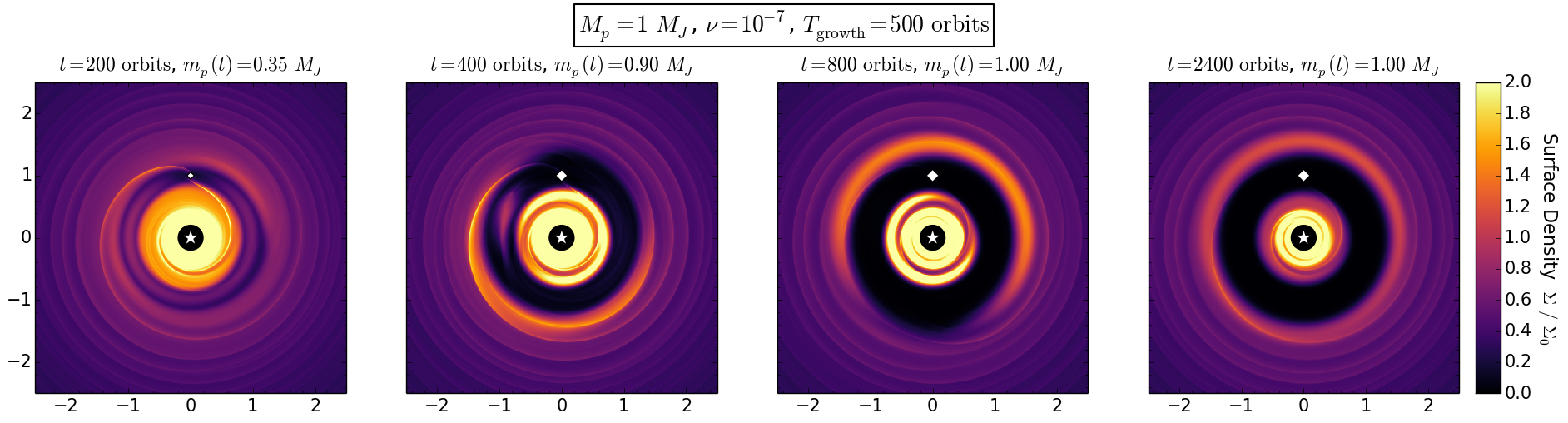}
\vspace*{2em}


\includegraphics[width=0.98\textwidth]{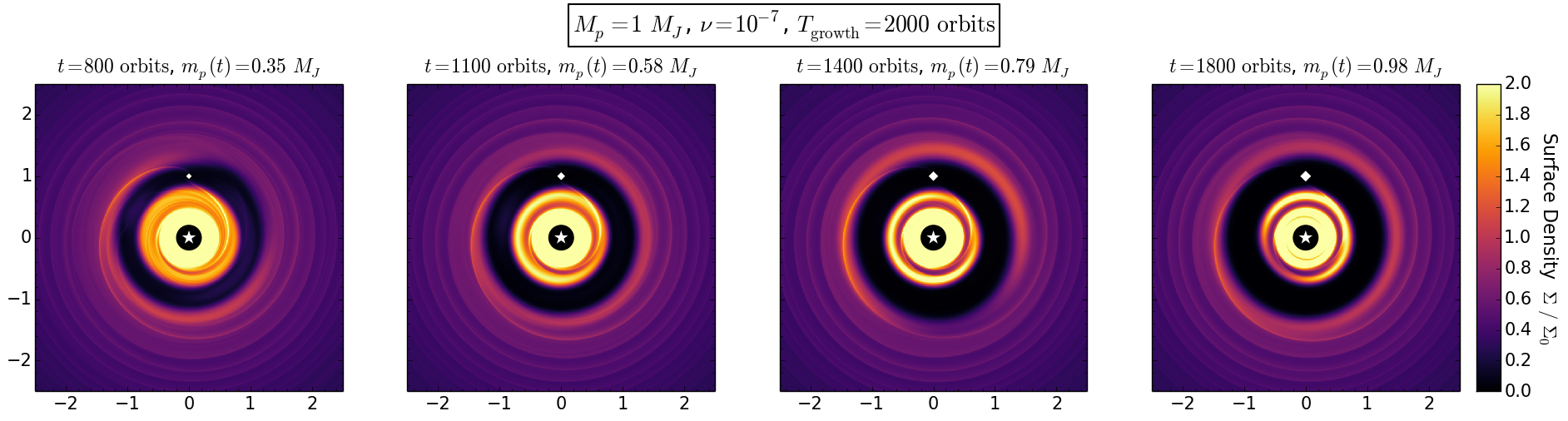}

\caption{Surface density snapshots for $M_p = 5~M_J$ and $\nu = 10^{-7}$, for $\Tg =$ 10,
  500, and 2000 $T_p$. See Figure~\ref{fig:pretty_m5_v7} for more
  details. 
 Top: With $T_{\rm growth} = 10~T_p$, a strong vortex forms by 100
 orbits, and survives for several thousand orbits. 
Middle: With $T_{\rm growth} = 500~T_p$, the planet triggers the
instability after 200 orbits, but the resulting $m~=~1$ vortex is much
more elongated with $\left[\Delta \phi\right]_\mathrm{min} \approx
240^{\circ}$. 
Bottom: With $T_{\rm growth} = 2000~T_p$, although a vortex forms by
800 orbits, it fades into a ring by 1100 orbits. Then, while the
planet is still growing, it re-triggers a vortex at 1400 orbits. This
new vortex has roughly the same density as the earlier one and like
the first, it also quickly dies into a ring just a few hundred orbits 
later.} 
\label{fig:pretty_m1_v7}
\end{figure*}

In the rapid growth case where $T_\mathrm{growth}=10~T_p$, we find that Jupiter-mass planets in
discs with a low viscosity ($\nu = 10^{-7}$) also induce strong,
concentrated vortices. These vortices have relatively long lifetimes of several thousand orbits, which is
consistent with previous studies \citep[e.g.][]{loboGomes15,
  les15} and only about a factor of $\sim3$ shorter than those with the higher mass planet. However for $T_\mathrm{growth} \ge 500~T_p$, lifetimes decrease by
an order of magnitude, azimuthal extents double ($\left[\Delta
  \phi\right]_\mathrm{min} \approx 240^{\circ}$ compared to
$120^{\circ}$) and $\Sigma_\mathrm{peak}$ drops very quickly to $<2~\Sigma_0$ (see
Figure~\ref{fig:comparisons}). Figure~\ref{fig:pretty_m1_v7} shows
surface density snapshots that highlight the  morphological
differences between these concentrated and elongated vortices over the
course of their lifetimes.

While slow growth does not suppress vortex formation entirely in low
viscosity discs, we find that in the higher viscosity disc with $\nu = 10^{-6}$, growth times over 1000 orbits inhibit vortex 
formation completely for $M_p = 1~\Mjup$. With 
$T_\mathrm{growth}$ of 250 and 500 $T_p$, planets create vortices
that are a little more elongated ($\left[\Delta
  \phi\right]_\mathrm{min} \approx 240^{\circ}$ compared to
$180^{\circ}$), have $\Sigma_\mathrm{peak}< 2~\Sigma_0$, and shorter
lifetimes by $\sim50\%$ compared to the rapid growth case. In these
slowly-grown cases, about half of the reported vortex lifetime
transpires before the high-$m$ RWI modes have merged into a single
vortex. For this reason along with their shorter lifetimes, vortices
generated by a Jupiter-mass planet in a high-viscosity disc with $\nu=10^{-6}$ are not
well-suited to explain observed disc asymmetries.

\subsection{Vortex Reformation} \label{ssec:reformation}
In a few cases, we observe a distinct behavior with large values of
$T_\mathrm{growth}$ (slow growth times). When a vortex is initially
weak, it can fade into a ring before the planet reaches its final
mass. Since the planet continues to clear out the gap as it grows,
these rings can develop sufficiently steep edges to re-trigger the RWI
and form a new vortex. This only occurs for weak vortices, which can
fade quickly before the growth phase ends and do not smooth out the
nearby gap edge as much as stronger vortices formed by quickly
accreting planets. Note that secondary vortices have also been
observed by \cite{loboGomes15} for discs with a non-isothermal
equation of state, albeit at a different location further from the
planet. 

The bottom panel of Figure~\ref{fig:pretty_m1_v7} shows an example of
a vortex reforming in a low viscosity disc ($\nu = 10^{-7}$) for a
$1~M_J$ planet with $T_{\rm growth} = 2000~T_p$. The  vortex initially
forms and dies out after several hundred orbits. Shortly after, it
reforms at a similar level of strength in terms of $\Sigma_{peak}$ and
$\left[\Delta \phi\right]_\mathrm{min}$, and also survives for several
hundred orbits. 

Along the same lines, we find the three cases of (A) $M_p = 5~M_J$,
$\nu = 10^{-7}$, $T_{\rm growth} = 4000~T_p$; (B) $M_p = 5~M_J$, $\nu
= 10^{-6}$, $T_{\rm growth} = 1000~T_p$; and (C) $M_p = 1~M_J$, $\nu =
10^{-6}$, $T_{\rm growth} = 500~T_p$ also create vortices that die out
and reform while the planet is still growing. The first two cases are
shown in the bottom panels of Figure~\ref{fig:pretty_m5_v7} and
Figure~\ref{fig:pretty_m5_v6}. The first case is unique because the
first-generation vortex dies out very early on in the planet's growth
phase. This allows the second-generation vortex to become much
stronger than the first one. 

We cannot determine if the stronger, concentrated vortices in our
study can reform, since they die out long after the planet has grown
to its full mass. However, all of the weaker vortices that reform
begin their second-generation phase at a level of strength near the
strength of the first-generation vortex shortly before it
dissipated. Therefore, even if the strongest vortices can reform, we
do not expect them to reach the densities or lifetimes of the
first-generation vortices unless the planet has a significant fraction
of its growth phase remaining.  

\subsection{Convergence Tests} \label{ssec:resolution}

The standard resolution in our parameter study resolves the disc
rather well in the vicinity of the planet and the vortex compared to
other studies of planet-induced vortices \citep[e.g][]{loboGomes15,
  les15, bae16}. Nonetheless, computational planet-disc interaction
studies are notoriously susceptible to resolution artifacts
\citep[e.g.][]{deValBorro07, munoz14}.  

To confirm the above trends that we observe, we ran convergence tests
for several cases: (1) $M_p = 5~M_J$, $\nu=10^{-6}$,
with $T_{\rm growth}=10$, $500$, and $1000~T_p$; and (2) $M_p = 1~M_J$, $\nu=10^{-7}$, with $T_{\rm
  growth}=500$ and 1000 $T_p$. We find that the
trend that increasing $\Tg$ produces increasingly weaker vortices is
upheld at each resolution. Therefore, we are confident that this trend
is physical.  

The tests for the $5~M_J$ planets converge well quantitatively,
producing vortex lifetimes and $\Sigma_\mathrm{peak}$ consistent with
the standard resolution simulations to within $15\%$. For $1~M_J$
planets, the relation between $\Tg$ and vortex lifetime remains, but
the lifetimes vary with resolution. Specifically, the lifetimes for
all cases decrease with increasing resolution, suggesting that the
estimated lifetimes we present in Figure~\ref{fig:comparisons} as well
as in Section \ref{ssec:constraints} may be upper limits. We suspect
that the difference in behavior as a function of mass is related to
the better resolution of the planet's Hill radius and spiral shocks
from the planets in the higher mass case, as vortex dissipation may be
related to its interactions with these planetary shocks \citep{fu14a}.  

\begin{figure*} 
\centering
\includegraphics[width=0.98\textwidth]{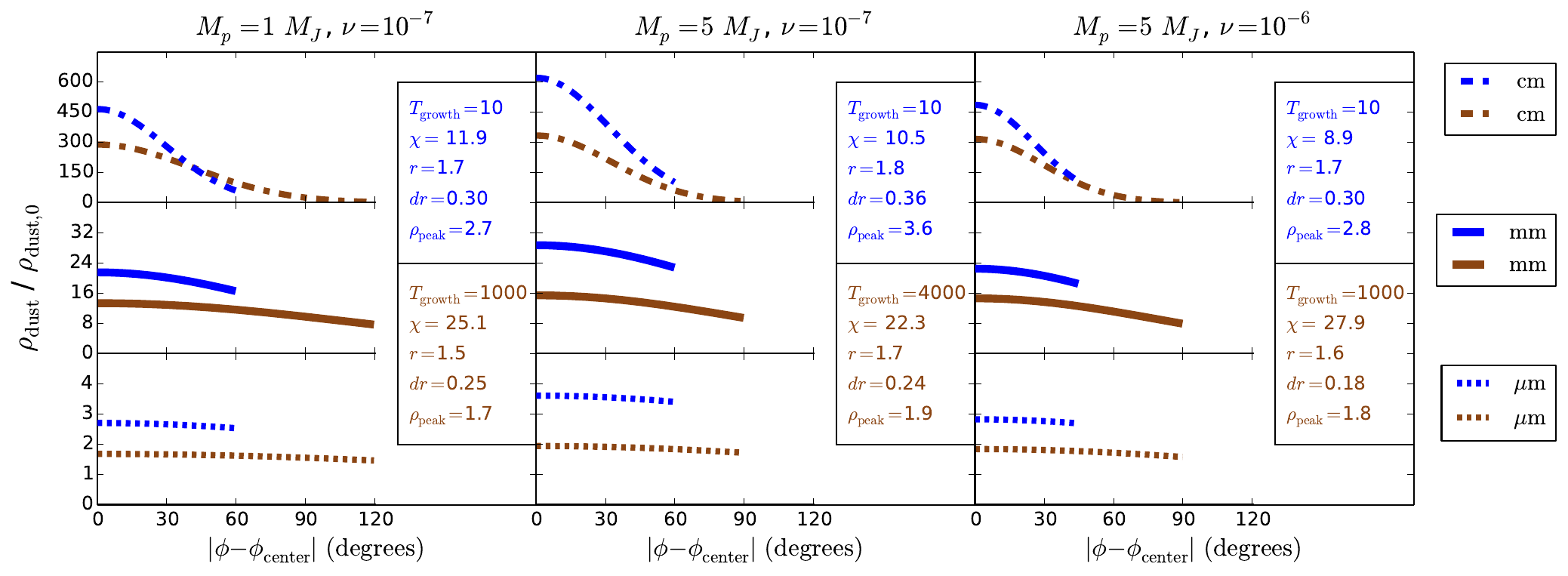}
\caption{Comparison of the dust distributions (cm-size, mm-size, and
  $\rm{\mu}$m-size) produced by vortices in three of the rapid growth
  cases (blue lines; parameters listed in the top windows) to those
  from a longer growth timescale ({brown} lines; parameters listed in
  the bottom windows). Distributions are calculated using
  Equation~\ref{eqn:dust}. The gas azimuthal extents can be extracted
  from these azimuthal dust density distributions through the dust
  extent of the $\rm{mm}$-size or $\rm{\mu m}$-size particles, or the
  steepness of the distribution of $\rm{cm}$-size particles. The
  distributions are normalized to the initial dust density at the
  location of the planet.
}
\label{fig:dust}
\end{figure*}



\section{Observational Implications} \label{sec:observations}

The disc asymmetries attributed to vortices are observed in mm-cm dust
grains \citep[e.g.][]{vanDerMarel13}. Although the gas overdensity may
be modest, a vortex acts as a very efficient dust trap due to the
local pressure maxima at the centre. To quantify the impact of planet
growth time on the properties of observed vortices, we must model the
distribution of dust particles in a range of sizes within our
simulated vortices. We have shown that $\Tg$ dramatically alters the
concentration and azimuthal extent of planet-induced vortices. To
illustrate the differences between concentrated and elongated vortices
in dust observations, we make use of the vortex dust-trapping model
developed by \cite{lyra13}. 


\subsection{Dust-Trapping Model} \label{ssec:dust}


We approximate our simulated vortices as Kida vortices with elliptical
streamlines and density contours of constant aspect ratio $\chi$
\citep{kida81}. We estimate the vortex aspect ratio as 
\begin{equation} \label{eqn:aspect} 
\chi = r \Delta \phi / \Delta r
\end{equation}
where $r\Delta \phi$ and $\Delta r$ are the vortex azimuthal and 
radial extents respectively, that are measured from our simulations.   
{The azimuthal extent is chosen to be $\left[\Delta
  \phi\right]_\mathrm{min}$ and the radial extent is estimated from the differenced surface density.} 

We then apply \citeauthor{lyra13}'s model for the steady-state dust
distribution in a disc vortex:  
\begin{equation} \label{eqn:dust} 
\rho_\mathrm{dust}(a) = \epsilon \rho_\mathrm{peak} (S + 1)^{3/2} \exp
\bigg\{-\frac{a^2 f^2(\chi)}{2H^2}(S + 1)\bigg\} 
\end{equation}
(see their Equation 64), where $a$ is the semi-minor axis of an
elliptical density contour inside the vortex; $\rho_\mathrm{peak}$ is
the peak gas density, here inferred from our simulations (see
Figure~\ref{fig:comparisons}); $S$ is the ratio of the particles'
Stokes number to the dimensionless diffusion coefficient in the
vortex; $f(\chi)$ is a scale function of order unity (see their
Equation 35); and $\epsilon$ is the dust-to-gas ratio. We fix $S$ to
be $3 \times 10^{-3}$, 3, and 30 to represent particles of  $\rm{\mu
  m}$-size, $\rm{mm}$-size, and $\rm{cm}$-size respectively. The value
of $S$ can vary by about an order of magnitude for each size for
reasonable choices of the diffusion coefficient.  

\subsection{Dust distributions in Concentrated and Elongated Vortices} \label{ssec:comparisons}

Figure~\ref{fig:dust} compares the $\rm{\mu m}$-size, $\rm{mm}$-size,
and $\rm{cm}$-size dust distributions of concentrated and elongated
vortices for three different combinations of planet masses and disc
viscosities. 

In general, we find that the distributions of $\rm{mm}$-size particles
are the best tracer for the underlying gas vortex's azimuthal extent
since they are effectively concentrated and maintain a fairly constant
overdensity from the vortex centre to the outer edge. 
The distributions of $\rm{\mu m}$-size particles are also relatively
flat throughout the vortex, but they are not trapped as efficiently
and thus have a lower density enhancement throughout. 

With larger particles ($\rm{cm}$-size), the dust extents cannot be
used to measure the gas extents directly since the dust concentrations
approach zero inside the outer edge of the vortex. Instead, we find
that the slopes of these steeper azimuthal distributions are a better
indicator of the gas azimuthal extent. However, the differences in
these slopes for vortices with different azimuthal extents can be very
small ($< 10 \%$) when the distributions are normalized to the density
at the vortex centre, making the extents difficult to measure
precisely with this size particle. 

This simplified model illustrates that two types of vortices might be
distinguishable at mm wavelengths. {Since $\rm{\mu m}$-sized particles are also good at tracing the azimuthal extents, any contamination to mm observations from smaller wavelengths should only help distinguish an elongated vortex from a concentrated vortex.} We expect that the dust
overdensities shown in Figure \ref{fig:dust} are likely underestimates
of the observable asymmetry, as we have not accounted for the removal
of dust elsewhere along the azimuth at the radial location of the
vortex. However, the simulated vortices have more complicated
streamline and density structures at their outer edges, which might
make dust trapping less effective.  A more detailed dust-trapping
model that is more appropriate for elongated vortices is needed to
better assess the viability of measuring a vortex's azimuthal extent
for constraining the associated planet's growth timescale.   



\subsection{Constraints from Planet Formation Models} \label{ssec:constraints}

We have shown that planets must form within a few thousand orbits or
less in order to induce long-lived vortices. We now compare these timescales with
realistic growth rates from planet formation models, finding that
extreme conditions are  required to satisfy the growth rate
constraint.  

First, we consider the fastest possible accretion rates. A strong
upper limit on the planet's growth is set by allowing the planet to
efficiently accrete from its feeding zone in the 2D limit as 
\begin{equation}
\Mdot_p \approx \Sigma \Omega_p R_H^2
\end{equation}
\citep{youdin13}. To evaluate this rate, we assume a surface density profile that is
proportional to the Minimum Mass Solar Nebula (MMSN),
\begin{equation}
\Sigma (r) = 2200~F
\left(\frac{r}{\mathrm{1~AU}}\right)^{-3/2} ~\mathrm{g}\,\mathrm{cm}^{-2} , 
\label{eq:mmsn}
\end{equation}
where $F$ is an enhancement factor ($F = 1$ refers to the MMSN). With 
this profile, the maximum growth rate of the planet is 
\begin{equation}
  \Mdot_p \approx 1.1 \times 10^{-2}~F  
  \left(\frac{r}{\mathrm{5~AU}}\right)^{1/2}\left(\frac{q}{\mathrm{5
      \times 10^{-4}}}\right)^{2/3}\frac{M_J}{T_p}, 
\end{equation}
where we have used the value of $q=0.5~M_J / M_{\odot}$ for the Hill radius scaling as the
planet cannot initially accrete from the Hill radius corresponding to its final mass.
This upper limit corresponds to $T_\mathrm{growth} =
M_p/\dot{M}_p \approx 90~T_p$ for a Jupiter-mass planet that forms at
5 AU in the MMSN. This timescale is safely an order of magnitude shorter than the critical growth rate 
required for vortex formation in our models ($T_p \sim 
1000~T_p$). Thus, if planets were able to accrete at this maximal
rate, vortex formation would be robust. However, achieving these
maximum growth rates is unlikely when a planet has already opened up a
gap, which occurs at a relatively low mass in a low viscosity disc.  

A second simple estimate for planet growth time can be derived from
considering the background accretion rate through the disc. In 
steady state, we expect the disc to funnel material into the vicinity
of the planet at {the} disc accretion rate $\dot{M}_d = 3\pi \nu
\Sigma$.
Since we expect $\dot{M}_p \lesssim \dot{M}_d$ {(\cite{lubow06} find that $\dot{M}_p$ never exceeds $80\%$ of $\dot{M}_d$)}, for low viscosity
discs the background accretion rate sets a more stringent limit on 
the planet growth times than the estimate above. Equating the disc
accretion rate to the planet growth rate, we define the critical
surface density for vortex formation to be 
\begin{equation}
\Sigma_{\rm crit} \equiv \frac{\dot{M}_{d}}{3 \pi \nu} = \frac{M_p}{T_\mathrm{growth,c}}\frac{1}{3\pi\nu},
\end{equation}
where $T_\mathrm{growth,c}$ is the critical planet growth {timescale} that
{still allows vortices to form} (which is a function of $M_p$ and $\nu$). As
above, we scale $\Sigma_{\rm crit}$ to the MMSN to obtain $F_{\rm
  crit}$, the enhancement over the MMSN required to allow sufficiently
rapid accretion to induce vortex formation. We find 
\begin{equation}
F_\mathrm{crit} = 29.1 \left(\frac{M_p}{1~M_J}\right)
\left(\frac{1000~T_p}{T_\mathrm{growth,c}}\right)\left(\frac{r}{5~\mathrm{AU}}\right)^{-1/2} 
\left(\frac{10^{-6}~r_p^2 \Omega}{\nu}\right),  
\label{eq:f_crit} 
\end{equation}
Aside from the discrepancy with observed disc masses \citep{andrews11},
this very high surface density is incompatible with the assumption of
a low disc viscosity, because it corresponds to a disc with a small
Toomre $Q$. For the passive irradiated disc profile from \cite{chiang97} (see their Eq. 14a) with
\begin{equation} \label{eq:t_profile}
T = T_\mathrm{mid}\left(\frac{r}{5~\mathrm{AU}}\right)^{-3/7},
 \end{equation}
 where $T_\mathrm{mid}$ is midplane temperature at 5 AU in the fiducial T-Tauri irradiation model,
\begin{equation} \label{eq:q_crit}
  Q  \approx 0.77
  \left(\frac{r}{5 ~\rm{AU}}\right)^{-\frac{3}{14}} \left(\frac{M_*}{M_\odot}\right)^{1/2} \left(\frac{T_\mathrm{mid}}{75~\rm{K}}\right)^{1/2} \left(\frac{29.1}{F_{\rm crit}}\right).
 \end{equation}
Discs with $Q < 2$ are
subject to gravitational instabilities that generate effective $\alpha
> 0.01$ or $\nu > 10^{-4}~r_p^2\Omega_p$ \citep{kratter16}. Even for a
factor of $\sim 5-10$ decrease in $F_\mathrm{crit}$ (increase in $Q$),
disc self-gravity may suppress large-scale vortex formation via the
RWI \citep{mkl11}. We conclude that the planet growth 
rates required for vortex formation are incompatible with growth via mass
re-supply from the disc.
  
Neither of the above estimates captures the details of planetary
accretion.  More realistic simulations of Jupiter's formation that
account for gap opening and feedback produce accretion rates in line
with our second estimate, which is 1 to 2 orders of magnitude slower
than the rates needed for robust vortex formation. \cite{lissauer09}
model the growth of a $1~M_J$ planet at 5~AU in a disc with $\nu =
10^{-6}r_p\Omega_p^2$ and $F \approx  3.5$. They find that the runaway
gas accretion phase lasts for $4 \times 10^4~T_p$ in the simulation
they deem the most realistic. Neglecting disc dispersal, they estimate
growth times of  $8000~T_p$. At even this faster rate, our simulations
predict that vortex formation would be inhibited. More detailed models
of accretion through circumplanetary discs produce similarly low
accretion rates, and thus slow growth times. Again, for planet
properties akin to Jupiter, studies find $\dot{M_p} \approx 10^{-5} 
M_J\,\mathrm{yr}^{-1}$ \citep[e.g.][]{papaloizou05, ayliffe09, rivier12}. If 
circumplanetary disc accretion rates were only weakly dependent on 
planet semi-major axis, this would favour vortex formation at larger
disc radii. 

Although our calculations above suggest that realistic planet growth times are slow, vortex formation may still be possible with less idealized disc conditions. For example, secondary
effects such as the efficient heating of material in the vicinity of an accreting planet (which we neglect in our isothermal discs) has been shown to help vortex formation and survival \citep{les15, loboGomes15, owen16b}. Even if the inclusion of more rigorous thermodynamics could ameliorate the inhibition of the instability caused by slow planet growth times, vortices induced by slowly growing giant planets should still be subject to the vortex back-reaction effects that lower their overdensities, widen their azimuthal extents, and shorten their lifetimes as in our study. As discussed in Section~\ref{ssec:comparisons}, these characteristics each likely severely damp prospects for observing vortices.


\section{Conclusions} \label{sec:conclusions}

In this paper, we explore the impact of a planet's growth timescale on
the properties of vortices induced at the outer gap edge through the
Rossby Wave Instability. We find that the values of $\Tg$ needed for
the formation of long-lived planet-induced vortices are short compared
to those expected from 
core accretion models or mass resupply in a low-viscosity
disc. Even with modest values of $\Tg \sim 1000~T_p$, the resulting vortices
have lower gas densities, more elongated azimuthal extents, and
shorter lifetimes that are less than half their survival times with
rapid planet growth. This is in contrast to vortex lifetimes on the
order of $10^3$ to $10^4~T_p$ obtained with $\Tg \lesssim 100~T_p$ in
previous studies \citep[e.g][]{fu14a}. 

With increased planet growth timescales, we find the RWI is in 
fact triggered soon after the planet mass reaches the gap-opening
criterion from \cite{crida06}, which is well below the final planet
masses studied here. Although the vortex evolution at the trigger
point is not sensitive to $\Tg$, the planet growth timescale strongly
affects the saturated state of the vortex. We attribute the transition to weaker
vortex properties to the competition between the timescale on which
the vortex smooths the gap edge as compared to the timescale on which
the growing planet steepens it. We also map out this transition as a
function of the dimensionless planet mass growth rate $\dot{q}$, and a
dimensionless measure of gap-edge steepness $q^2\mathrm{Re}$, where
$\mathrm{Re}$ is the Reynolds number \citep{varniere04}. If the
initial vortex forms when the planet is relatively low in mass, the
vortex back-reaction on the gap structure is more pronounced, stunting
any further growth of the vortex during subsequent planet
growth. Conversely, a short $\Tg$ allows the planet to sculpt the gap
edge much faster than the vortex can back-react on it.

A unique feature of the vortices induced by slowly-grown planets are
their much wider azimuthal extents, which can be twice that of
vortices induced by rapidly-grown planets. We caution that these
elongated vortices (in particular, those induced by lower mass
planets) may not survive as long as in our standard resolution
simulations, since their lifetimes are halved in our higher resolution
simulations. 

If such elongated vortices survive long enough to be
observable, we can measure their azimuthal extents from dust
observations. 
Since a planet's growth timescale preferentially affects the azimuthal
extents of the vortex it creates, a measurement of a vortex's
azimuthal extent could be used in conjunction with other properties
(such as the planet's mass, along with the disc's viscosity and
temperature) to provide limits on how quickly the associated planet
formed. Other effects that weaken vortices and cause them to be
shorter-lived (such as sub-optimal disc temperatures and viscosities, 
and dust feedback) do not affect their extents nearly as much as the
planet's growth rate. 

A low-viscosity disc is generally required for planet-induced vortex
formation \citep[e.g.][]{deValBorro07, fu14a}. However, the disc
viscosity also limits the accretion rate onto the planet, thereby
increasing its growth timescale. In this regime, the corresponding values of $\Tg$ prevent a planet from inducing a long-lived vortex. If the asymmetric discs observed with
ALMA are induced by vortices from giant planets, more detailed thermodynamics may need to be incorporated into 
future simulations to model them properly. Accounting for the disc cooling timescale or planetary accretion luminosity
may lead to stronger vortices even with slow growth rates \citep{les15, loboGomes15,owen16b}


Finally, we note that vortices associated with dead zone boundaries
\citep[e.g.][]{regaly12} are also expected to be elongated. As a result, one would
not be able to distinguish the origin of elongated vortices (as induced by a dead zone
boundary or a slowly-grown planet) based on azimuthal extent alone.

\section*{Acknowledgements}
{We thank Ruobing Dong, James Owen, Sijme-Jan Paardekooper, Nienke van der Marel, and the anonymous referee for their helpful comments.} MH is supported by the NSF Graduate Research Fellowship under Grant No. DGE 1143953. KMK is supported by the National Science Foundation
under Grant No. AST-1410174. MKL is supported by the Steward Theory Fellowship. The El 
Gato supercomputer, which is supported by the National
Science Foundation under Grant No. 1228509, was used to run all of the simulations in this study.


\bibliography{vortex_bibliography}

 \end{document}